\documentclass[twocolumn,letter]{jpsj3}
\usepackage{txfonts}
\usepackage{color}
\usepackage{ulem}

\title{
Superconductivity in Hexagonal BaPtAs: SrPtSb- and YPtAs-type Structures with Ordered Honeycomb Network
}

\author{
Kazutaka Kudo$^1$\thanks{kudo@science.okayama-u.ac.jp}, Takaaki Takeuchi$^1$, 
Hiromi Ota$^2$, 
Yuki Saito$^1$, Shin-ya Ayukawa$^1$, 
\\ Kazunori Fujimura$^1$, 
and Minoru Nohara$^1$\thanks{nohara@science.okayama-u.ac.jp}
}

\inst{
$^1$Research Institute for Interdisciplinary Science, Okayama University, Okayama 700-8530, Japan
\\ 
$^2$Advanced Science Research Center, Okayama University, Okayama 700-8530, Japan
}

\abst{
The crystal structure and superconductivity of hexagonal BaPtAs are reported. 
Single-crystal X-ray diffraction, magnetization, electrical resistivity, and specific heat measurements were performed in this study. 
Two hexagonal structures with different PtAs honeycomb network stacking sequences, namely, SrPtSb- (space group $P\bar{6}m2$, $D_{3h}^1$, No.~187) and YPtAs-type ($P6_3/mmc$, $D_{6h}^4$, No.~194) structures, were identified and found to exhibit superconductivity at 2.8 and 2.1--3.0 K, respectively. 
In contrast, the cubic LaIrSi-type structure ($P2_13$, $T^4$, No. 198) did not exhibit superconductivity above 0.1 K. 
BaPtAs provides a unique opportunity to study superconductivity with broken and preserved spatial inversion symmetry.
}

\begin{document}
\maketitle

Alkaline-earth platinum pnictides AEPtPn (AE = Ca, Sr, Ba; Pn = P, As, Sb) exhibit a variety of hexagonal structures that are characterized by PtPn honeycomb networks \cite{Wenski_Ca,Wenski_SrBa}. 
For instance, CaPt$_{x}$P$_{2-x}$\cite{Wenski_Ca} and SrPt$_{x}$P$_{2-x}$\cite{Wenski_SrBa} crystallize in an AlB$_2$-type structure ($P6/mmm$, $D_{6h}^1$, No. 191), in which the Ca(Sr) occupies the Al site, while the Pt and P are statistically distributed at the B site of the honeycomb layers, as shown in Fig. 1(a). 
SrPtSb and BaPtSb crystallize in a SrPtSb-type structure ($P\bar{6}m2$, $D_{3h}^1$, No. 187)\cite{Wenski_SrBa}, which is a ternary ordered variant of the AlB$_2$-type structure: The Pt and Sb atoms alternately occupy the B site, forming a PtSb ordered honeycomb network. 
The honeycomb layers are stacked along the $c$-axis in such a manner that a Pt and Sb atom is above each Pt and Sb atom, respectively, forming a --Pt(Sb)--Pt(Sb)-- stacking sequence, as shown in Fig. 1(b). 
Thus, the spatial inversion symmetry is globally broken, although the structure is neither polar nor chiral, because of the presence of mirror symmetries. 
SrPt$_x$(Pt$_{0.10}$As$_{0.9}$) and BaPt$_x$(Pt$_{0.10}$As$_{0.9}$) with a Pt defect of $0.60 \leq x \leq 0.75$ also crystallize in this structure type\cite{Wenski_SrBa}. 
In contrast, SrPtAs crystallizes in a KZnAs-type structure ($P6_3/mmc$, $D_{6h}^4$, No. 194),\cite{Wenski_SrBa} which is another ordered variant of the AlB$_2$-type structure: The PtAs ordered honeycomb layers are stacked along the $c$-axis in such a manner that an As and Pt atom is above each Pt and As atom, respectively, forming a --Pt(As)--As(Pt)-- stacking sequence, as shown in Fig. 1(c). 
Hence, the structure is globally centrosymmetric, although the spatial inversion symmetry is locally broken in the PtAs honeycomb network. 
In contrast, BaPtP and BaPtAs have been known to crystallize in the cubic LaIrSi-type structure ($P2_{1}3$, $T^4$, No. 198)\cite{Wenski_SrBa}, as shown in Fig. 1(d), which is the ternary ordered variant of the cubic SrSi$_2$-type structure ($P4_1$32, $O^7$, No. 213).

Among them, the hexagonal compounds with ordered honeycomb networks SrPtAs and BaPtSb exhibit superconductivity at 2.4 and 1.64 K, respectively\cite{Nishikubo,Kudo}.
The noncentrosymmetric structure of the ordered PtAs honeycomb network, strong spin-orbit coupling of Pt, and weak interlayer coupling between the PtAs layers make SrPtAs a unique medium to study theoretically predicted exotic superconductivity\cite{Goryo,Fischer,Wang_Sr,Youn,Akbari,Sigrist,Fischer2,Goryo2,Sumita}, such as the singlet-triplet mixed state\cite{Goryo}, chiral $d$-wave state\cite{Fischer}, and $f$-wave state\cite{Wang_Sr}.
These predictions have been experimentally examined on SrPtAs: The breaking of time-reversal symmetry in the superconducting state was observed by $\mu$SR measurements, suggesting chiral $d$-wave superconductivity as the most likely pairing state\cite{Biswas}, whereas a conventional $s$-wave pairing state was suggested by NMR/NQR\cite{Matano} and magnetic penetration depth\cite{Landaeta} measurements. 
No Hebel--Slichter coherence peak was observed and two-gap superconductivity was suggested by another NQR measurement\cite{Bruckner}. 
The superconducting state of BaPtSb has not yet been fully examined\cite{Kudo}. 
\begin{figure*}[t]
\begin{center}
\includegraphics[width=17cm]{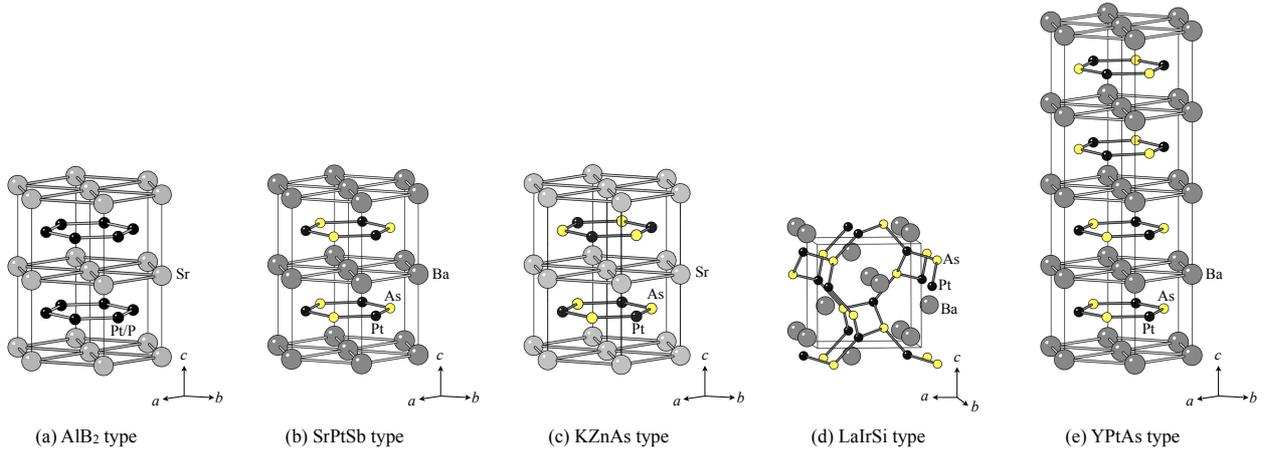}
\caption{
(Color online) Crystal structures of the (a) AlB$_2$-type SrPt$_x$P$_{2-x}$ ($P6/mmm$, $D_{6h}^1$, No. 191)\cite{Wenski_SrBa}, (b) SrPtSb-type BaPtAs ($P\bar{6}m2$, $D_{3h}^1$, No. 187), (c) KZnAs-type SrPtAs ($P6_3/mmc$, $D_{6h}^4$, No. 194)\cite{Wenski_SrBa}, (d) LaIrSi-type BaPtAs ($P2_13$, $T^4$, No. 198)\cite{Wenski_SrBa}, and (e) YPtAs-type BaPtAs ($P6_3/mmc$, $D_{6h}^4$, No. 194).
}
\end{center}
\end{figure*}

In this Letter, we report novel hexagonal structures of BaPtAs, namely, SrPtSb- ($P\bar{6}m2$, $D_{3h}^1$, No. 187) and YPtAs-type ($P6_3/mmc$, $D_{6h}^4$, No. 194) structures. 
Both structural phases exhibit superconductivity at 2.8 and 2.1--3.0 K, respectively. 
In contrast, the previously known cubic LaIrSi-type ($P2_13$, $T^4$, No. 198) phase does not exhibit superconductivity above 0.1 K\cite{rho_0.1K}.

Three structure types of BaPtAs were obtained by different heat treatments. 
First, stoichiometric amounts of Ba, Pt, and PtAs$_2$ powders were placed in an alumina crucible, sealed in an evacuated quartz tube, and heated at 1150 $^{\circ}$C for 24--72 h. 
The subsequent furnace cooling resulted in a hexagonal SrPtSb-type structure, while the rapid cooling in iced water resulted in a hexagonal YPtAs-type structure. 
For both cases, we obtained single crystals with typical dimensions of 0.2--0.5 $\times$ 0.2--0.5 $\times$ 0.2--0.5 mm$^{3}$. 
The YPtAs-type structure was obtained as a single phase, while the SrPtSb-type structure was obtained as a mixture with a powder of the cubic LaIrSi-type structure\cite{powderXRD}. 
In contrast, the heating at 1100 and 1000 $^{\circ}$C resulted in a cubic LaIrSi-type BaPtAs and defected SrPtSb-type BaPt$_x$(Pt$_{0.10}$As$_{0.9}$) as the main phases, respectively. 
The SrPtSb- and YPtAs-type BaPtAs samples decomposed in air within several days, while the LaIrSi-type BaPtAs and the defected SrPtSb-type BaPt$_x$(Pt$_{0.1}$As$_{0.9}$) samples were stable in air\cite{Wenski_SrBa}.

The obtained samples were characterized by powder X-ray diffraction (XRD) using a Rigaku MiniFlex600 X-ray diffractometer with Cu$K_{\alpha}$ radiation\cite{powderXRD} and by single-crystal XRD using a Rigaku single-crystal X-ray structural analyzer (Varimax with Saturn). 
Single-crystal XRD experiments were performed using small single crystals with typical dimensions of 0.04--0.06 $\times$ 0.03--0.05 $\times$ 0.01--0.05 mm$^3$. 
The magnetization $M$ was measured using the Magnetic Property Measurement System by Quantum Design. 
The electrical resistivity $\rho$ and specific heat $C$ were measured using the Physical Property Measurement System (PPMS) by Quantum Design.

The single-crystal structure analysis of BaPtAs samples obtained via different heat treatments revealed the presence of two hexagonal structures that have not been reported thus far, namely, SrPtSb- ($P\bar{6}m2$, $D_{3h}^1$, No. 187)\cite{SrPtSb} and YPtAs-type ($P6_3/mmc$, $D_{6h}^4$, No. 194) structures. The crystallographic data are summarized in Tables I and II. 
\begin{table}[t]
\caption{Data collection and refinement statistics for the single-crystal X-ray structure analysis of two polymorphs of BaPtAs at 100 K. The radiation type was Mo $K_\alpha$ with a wavelength of 0.71075 \AA. 
The crystal of YPtAs-type structure was picked up from the same batch as the sample \#1 shown in Figs. 3 and 4.
}
\begin{center}
\begin{tabular}{lcc}
\hline
Formula & BaPtAs & BaPtAs\\ 
Structure type   &  SrPtSb & YPtAs \\
Crystal system & hexagonal & hexagonal \\
Space group     & $P\bar{6}m2$ ($D_{3h}^1$, No. 187)        & $P6_3/mmc$ ($D_{6h}^4$, No. 194)\\
$a$ (\AA)        & 4.308(4) & 4.324(4) \\
$c$ (\AA)        & 4.761(5) & 19.101(8) \\
Volume (\AA$^3$)  & 76.52(13) & 309.2(4) \\
$Z$ value                    & 1 & 4 \\
$R1$ (\%) [$I > 2\sigma(I)$] & 4.66 & 2.67 \\
\hline
\end{tabular}
\end{center}
\label{default}
\end{table}
\begin{table}[t]
\caption{Occupancies, atomic coordinates, and equivalent isotropic atomic displacement parameters at 100 K listed for two hexagonal polymorphs of BaPtAs with the SrPtSb-type ($P\bar{6}m2$, $D_{3h}^1$, No. 187) and YPtAs-type ($P6_3/mmc$, $D_{6h}^4$, No. 194) structures. 
}
\begin{center}
\begin{tabular}{llllll}
\hline
Site & Occupancy & $x/a$ & $y/b$ & $z/c$ & 100$U_{\rm eq}$ (\AA$^2$) \\
\hline
\multicolumn{6}{c}{SrPtSb type}\\ 
Ba & 1 & 0 & 0 & 0 & 0.28(9)\\
Pt1/As2 & 0.961(13)  & 1/3 & 2/3 & 1/2 & 0.25(7)\\
                & /0.039(13) & & & \\
As1/Pt2 & 0.961(13) & 2/3 & 1/3 & 1/2 & 0.09(11)\\
                & /0.039(13) & & & \\
\hline
\multicolumn{6}{c}{YPtAs type}\\
Ba1 & 1 & 0 & 0 & 0 & 0.11(5)\\
Ba2 & 1 & 0 & 0 & 1/4 & 0.61(6)\\
Pt & 1 & 1/3 & 2/3 & 0.12437(5) & 0.36(3)\\
As & 1 & 1/3 & 2/3 & 0.62640(14) & 0.40(5)\\
\hline
\end{tabular}
\end{center}
\label{default}
\end{table}

The SrPtSb-type structure of BaPtAs consists of alternating stacked Ba triangular and ordered planar PtAs honeycomb layers, as shown in Fig. 1(b). 
The structure can be formulated as Ba(Pt$_{1-\delta}$As$_{\delta}$)(Pt$_{\delta}$As$_{1-\delta}$) with a tiny amount of inter-site mixing $\delta$ = 0.039(13), where (Pt$_{1-\delta}$As$_{\delta}$) and (Pt$_{\delta}$As$_{1-\delta}$) occupy the positions of the honeycomb network alternately, forming an ordered honeycomb network, and the SrPtSb-type structure ($P\bar{6}m2$, $D_{3h}^1$, No. 187) is preserved. 
This contrasts with the isostructural SrPtSb and BaPtSb, in which no inter-site mixing between Pt and Sb has been suggested experimentally. \cite{Wenski_SrBa}

The YPtAs-type structure of BaPtAs consists of alternating stacked Ba triangular, ordered, and weakly-puckered PtAs honeycomb layers, as shown in Fig. 1(e). 
No inter-site mixing between Pt and As is suggested, as shown in Table II. 
This structure is the four-layer superstructure derivative of the AlB$_{2}$-type structure.
The PtAs ordered honeycomb layers are stacked along the $c$-axis in such a manner that forms a --As(Pt)--As(Pt)--Pt(As)--Pt(As)-- stacking sequence.
The position of adjacent As atoms in the As(Pt)--As(Pt) bilayer, as well as in the equivalent Pt(As)--Pt(As) bilayer, is shifted in the direction facing each other, while the position of adjacent Pt atoms is shifted in the opposite direction, resulting in the puckering of PtAs layers. 
The difference in the $z$ coordinates of Pt and As in the honeycomb network $\Delta z/c$ is estimated to be 0.002, which is much smaller than 0.016 for YPtAs, \cite{Wenski_Y} suggesting the very weak puckering in BaPtAs. 
Note that the structure is globally centrosymmetric, although the spatial inversion symmetry is locally broken in the PtAs honeycomb network, similar to SrPtAs\cite{Wenski_SrBa}.

The SrPtSb-type BaPtAs exhibits superconductivity at 2.8 K. 
As shown in Fig.~2(a), the temperature dependence of magnetization $M$ shows diamagnetic behavior below 2.8 K with a large shielding signal. 
The temperature dependence of resistivity $\rho$ exhibits a sharp drop below 2.82 K and becomes negligibly small below 2.75 K. 
The magnetic field dependence of $\rho$ demonstrates that the compound is a type-II superconductor, in which superconductivity persists up to a relatively high magnetic field. 
As shown in the inset of Fig. 2(b), the upper critical field $H_{\rm c2}$, which was determined from the midpoint of the resistive transition, increases almost linearly with the decrease in temperature down to 1.8 K. 
We estimated the value of $H_{\rm c2}$ at 0 K to be 5.5 kOe from the extrapolating line. The Ginzburg--Landau coherence length $\xi_0$ was estimated to be 245 \AA \ from $\xi_0 = [\Phi_0/2\pi H_{\rm c2}(0)]^{1/2}$, where $\Phi_0$ is the magnetic flux quantum. 
\begin{figure}[t]
\begin{center}
\includegraphics[width=6.5cm]{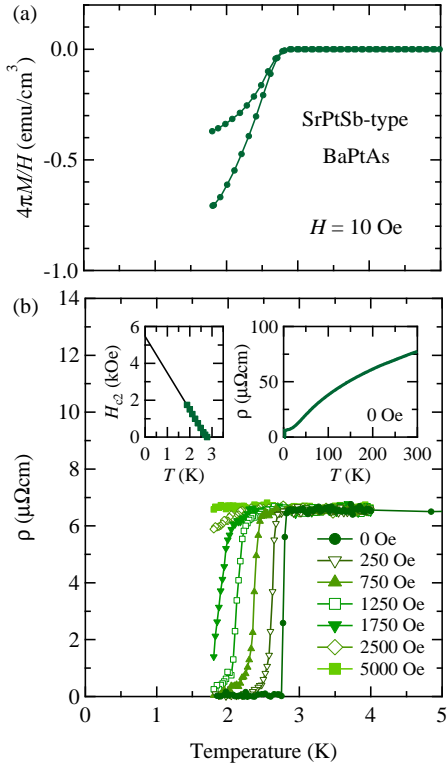}
\caption{
(Color online) Temperature dependences of (a) magnetization divided by magnetic field, $M/H$, under field and zero-field cooling conditions and (b) electrical resistivity $\rho$ at different magnetic fields for SrPtSb-type BaPtAs. 
The  $\rho$ measurements were performed using a multigrain crystal mixed with the SrPtSb- and LaIrSi-type BaPtAs. 
The left inset of (b) presents the temperature dependence of the upper critical field $H_{\rm c2}$. The solid line is an eye guide. The right inset of (b) shows the electrical resistivity dependent on temperature in a wide temperature range. 
}
\end{center}
\end{figure}
\begin{figure}[t]
\begin{center}
\includegraphics[width=6.5cm]{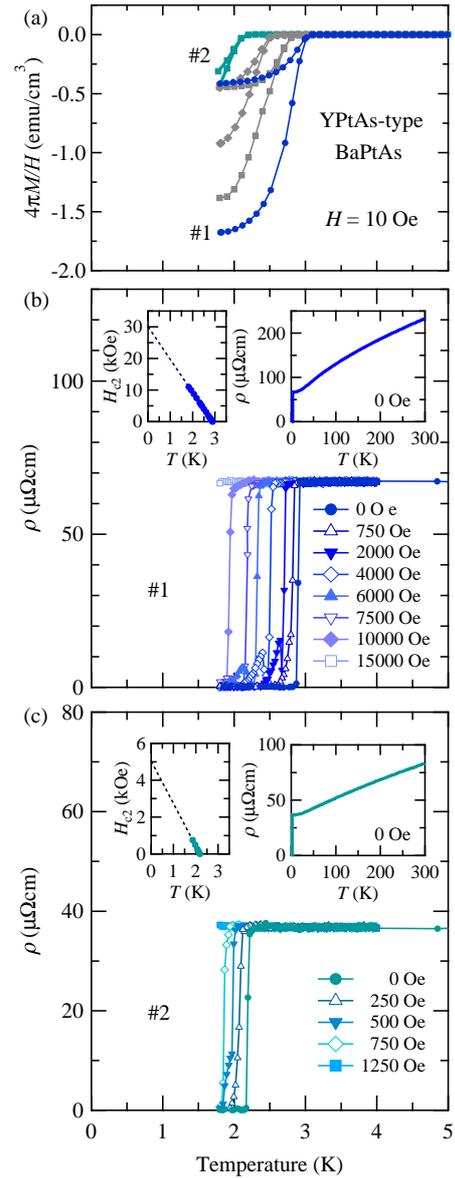}
\caption{
(Color online) Temperature dependences of (a) magnetization divided by magnetic field, $M/H$, under field and zero-field cooling conditions, and (b)(c) electrical resistivity $\rho$ in various magnetic fields for the different batch single-cysralline samples of the YPtAs-type BaPtAs. 
The left insets of (b) and (c) present the temperature dependence of the upper critical field $H_{\rm c2}$. The solid lines are eye guides. The right insets of (b) and (c) show the electrical resistivity in a wide temperature range. 
The electrical resistivity $\rho$ was measured with a current parallel to the in-plane [1$\bar{2}$10] direction and magnetic fields perpendicular to the (10$\bar{1}\bar{1}$) cleavage surface. 
}
\end{center}
\end{figure}
\begin{figure}[ht]
\begin{center}
\includegraphics[width=6.5cm]{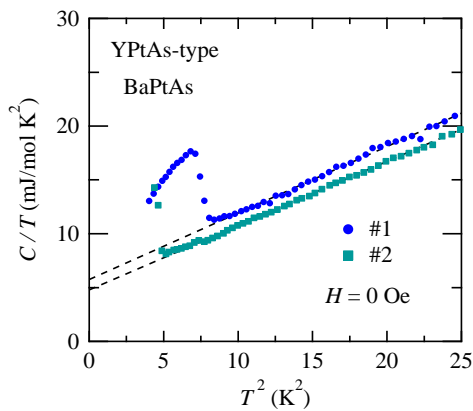}
\caption{
(Color online) Specific heat divided by temperature, $C/T$, vs $T^2$ plot for the different batch samples of the YPtAs-type BaPtAs. 
The dashed lines denote the fit by $C/T = \gamma + \beta T^{2}$, where $\gamma$ is the electronic specific heat coefficient and $\beta$ is a constant corresponding to the Debye phonon contributions.
}
\end{center}
\end{figure}

The YPtAs-type BaPtAs exhibits superconductivity at 2.1--3.0 K. 
To the best of our knowledge, this is the first Pt-based superconductor with a YPtAs-type structure. 
As shown in Fig. 3(a), the temperature dependence of magnetization $M$ exhibits diamagnetic behavior with a large shielding signal. 
The value of $T_{\rm c}$ is found to be distributed in the range of 2.1--3.0 K among the batches, although their lattice parameters are almost the same\cite{LatticeParams}. 
Figures 3(b) and 3(c) present the temperature dependence of resistivity $\rho$ for the typical samples. 
The resistivity for samples \#1 and \#2 exhibits a sharp drop below 2.92 and 2.22 K and becomes negligibly small below 2.87 and 2.17 K, respectively.
In both samples, the $T_{\rm c}$ gradually decreases with increasing magnetic field. 
As shown in the insets of Figs. 3(b) and 3(c), the $H_{\rm c2}$(0) of 30 and 5 kOe and the corresponding $\xi_0$ of 105 and 256 \AA \ are estimated for samples \#1 and \#2, respectively. 
The $H_{\rm c2}$(0) of sample \#1 was found to be one order of magnitude larger than that of sample \#2 and those of the other Pt-based honeycomb superconductors, SrPtAs ($T_{\rm c}$ = 2.4 K, $H_{\rm c2}$(0) = 2.2 kOe) \cite{Nishikubo} and the SrPtSb-type BaPtAs (2.8 K, 5.5 kOe).
The reason for the enhanced $H_{\rm c2}$ of \#1 is unclear at present and is an important issue to be clarified.

The resistivity data exhibit a sharp transition at $T_{\rm c}$ in the zero-magnetic field for both the SrPtSb- and YPtAs-type BaPtAs. 
However, for low fields, such as 500 Oe shown in Fig. 3(c), the resistivity data present a kink immediately below the onset temperature before exhibiting zero resistivity. 
For higher magnetic fields, such as 4000 Oe shown in Fig. 3(b), these anomalies in the resistivity data become more prominent, and the resistivity exhibits a sharp dip immediately below the onset temperature, followed by a sharp peak, and then gradually approaches zero with decreasing temperature. 
This is called the ``peak effect'' and is observed in some type-II superconductors. \cite{Kaluarachchi}
The underlying mechanism is not yet fully understood.

Figure 4 presents the temperature dependence of specific heat for the YPtAs-type BaPtAs. 
Measurements for the SrPtSb-type structure were not carried out, because of the small number of single crystals. 
A clear jump can be observed at $T_{\rm c}$ in both samples \#1 and \#2. 
The assumption of an ideal jump at $T_{\rm c}$ to satisfy the entropy conservation at the transition gives the estimates of $T_{\rm c} =$ 2.75 K and $\Delta C/T_{\rm c} =$ 8.40 mJ/K$^2$mol for sample \#1. 
The normal-state data could be fitted by $C/T = \gamma + \beta T^2$, where $\gamma$ is the electronic specific heat coefficient and $\beta$ is the coefficient of phonon contributions from which the Debye temperature $\Theta_{\rm D}$ is estimated, yielding the estimates of $\gamma =$ 5.74 and 4.77 mJ/K$^2$mol and $\Theta_{\rm D} =$ 211 and 213 for \#1 and \#2, respectively. 
The $\gamma$ value of \#1 is larger than that of \#2, while the values of $\Theta_{\rm D}$ are almost the same, suggesting that the difference in $T_{\rm c}$ between the different batches is most likely attributed to the difference in the density of states at the Fermi level. 
For sample \#1, the normalized jump $\Delta C/\gamma T_{\rm c}$ was estimated to be 1.46, which is almost coincident with the value of the BCS weak coupling superconductivity (1.43).

In conclusion, the equiatomic ternary compound BaPtAs was found to exhibit polymorphism.
In addition to the previously known cubic LaIrSi-type structure ($P2_13$, $T^4$, No. 198), we identified two new hexagonal structures, namely, SrPtSb- (space group $P\bar{6}m2$, $D_{3h}^1$, No.~187) and YPtAs-type structures ($P6_3/mmc$, $D_{6h}^4$, No.~194), using single-crystal XRD measurements. 
These structures are characterized by ordered PtAs honeycomb networks with a uniform --As(Pt)--As(Pt)-- stacking sequence and a four-layer superstructure of a --As(Pt)--As(Pt)--Pt(As)--Pt(As)-- stacking sequence, respectively.
Spatial inversion symmetry is globally broken in the former structure, whereas it is preserved in the latter. 
The magnetization, electrical resistivity, and specific heat measurements revealed that the SrPtSb- and YPtAs-type BaPtAs are bulk superconductors with $T_{\rm c}$ = 2.8 and 2.1--3.0 K, respectively. 
In contrast, the cubic LaIrSi-type structure ($P2_13$, $T^4$, No. 198) was found to not exhibit superconductivity above 0.1 K. 
The discovery of superconductivity in the hexagonal BaPtAs with a PtAs ordered honeycomb network provides opportunities not only for the experimental examination of theoretically predicted chiral $d$-wave superconductivity but also for further theoretical studies on exotic states that result from the strong spin-orbit interaction of Pt under broken inversion symmetry.

\begin{acknowledgments}
The authors are grateful to Seiichiro Onari and Jun Goryo for their valuable discussions, and to Naoki Nishimoto, Satoshi Ioka, and Hikaru Hiiragi for their help in the experiments. This work was partially supported by Grants-in-Aid for Scientific Research (15H05886 and 16K05451) provided by the Japan Society for the Promotion of Science (JSPS) and the Program for Advancing Strategic International Networks to Accelerate the Circulation of Talented Researchers (R2705) from JSPS. 
\end{acknowledgments}

\onecolumn
\normalsize
\renewcommand{\thefigure}{S\arabic{figure}}
\setcounter{figure}{0}

\begin{center}
\section*{SUPPLEMENTAL MATERIALS}
\end{center}

\vspace{1cm}

\subsection*{\bf 1. The temperature dependence of electrical resistivity $\rho$ for the LaIrSi-type BaPtAs}
\vspace{0.5cm}
As shown in Fig. S1, the cubic LaIrSi-type BaPtAs did not exhibit superconductivity above 0.1 K. 
\begin{figure}[h]
\begin{center}
\includegraphics[width=6.5cm]{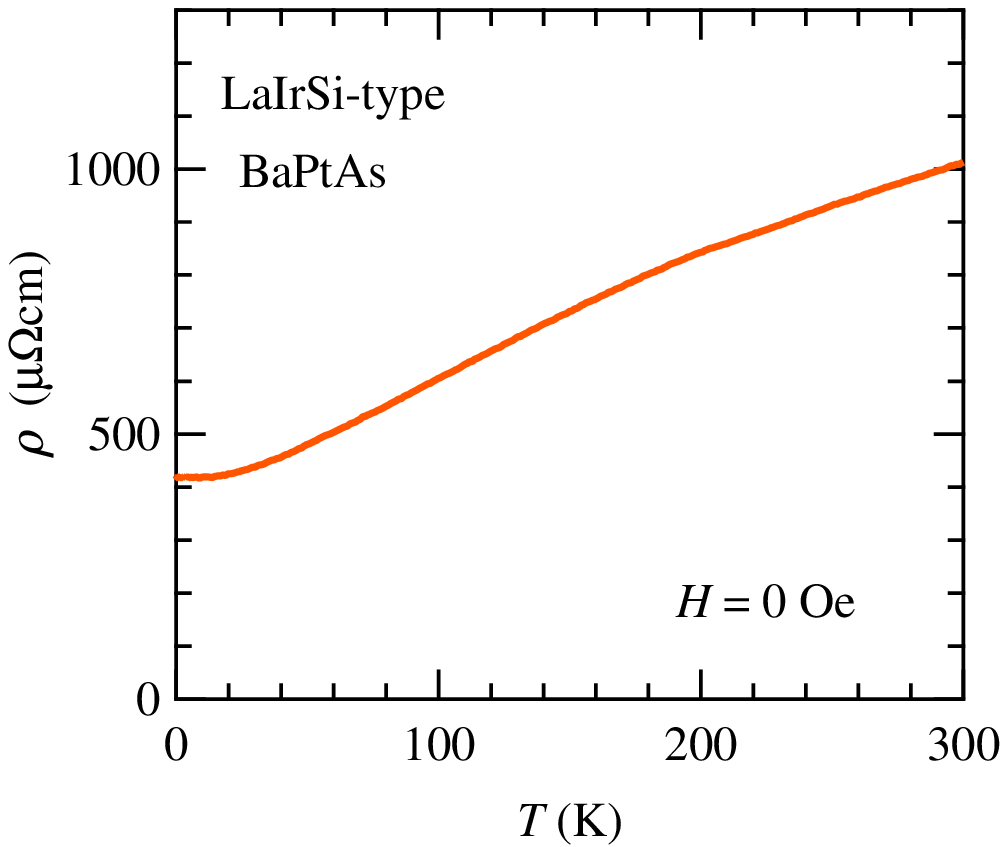}
\caption{
The temperature dependence of electrical resistivity $\rho$ down to 0.1 K for the cubic LaIrSi-type BaPtAs (space group $P2_13$, $T^4$, No. 198). 
The measurements were carried out using the AC Transport and Adiabatic Demagnetization Refrigerator options of the Physical Property Measurement System by Quantum Design. 
The frequency of excitation was 13 Hz and the amplitude of current density was 0.53 A/cm$^2$.  
}
\end{center}
\end{figure}

\clearpage
\subsection*{\bf 2. The powder X-ray diffraction (XRD) patterns for the SrPtSb- and YPtAs-type BaPtAs samples}
\vspace{0.5cm}
As shown in Fig. S2, the YPtAs-type phase was obtained as a single phase, while the SrPtSb-type phase was obtained as a mixture with a powder of the cubic LaIrSi-type phase. 
The YPtAs- and SrPtSb-type phases  were obtained without mixed with each other.   
\begin{figure}[h]
\begin{center}
\includegraphics[width=6.5cm]{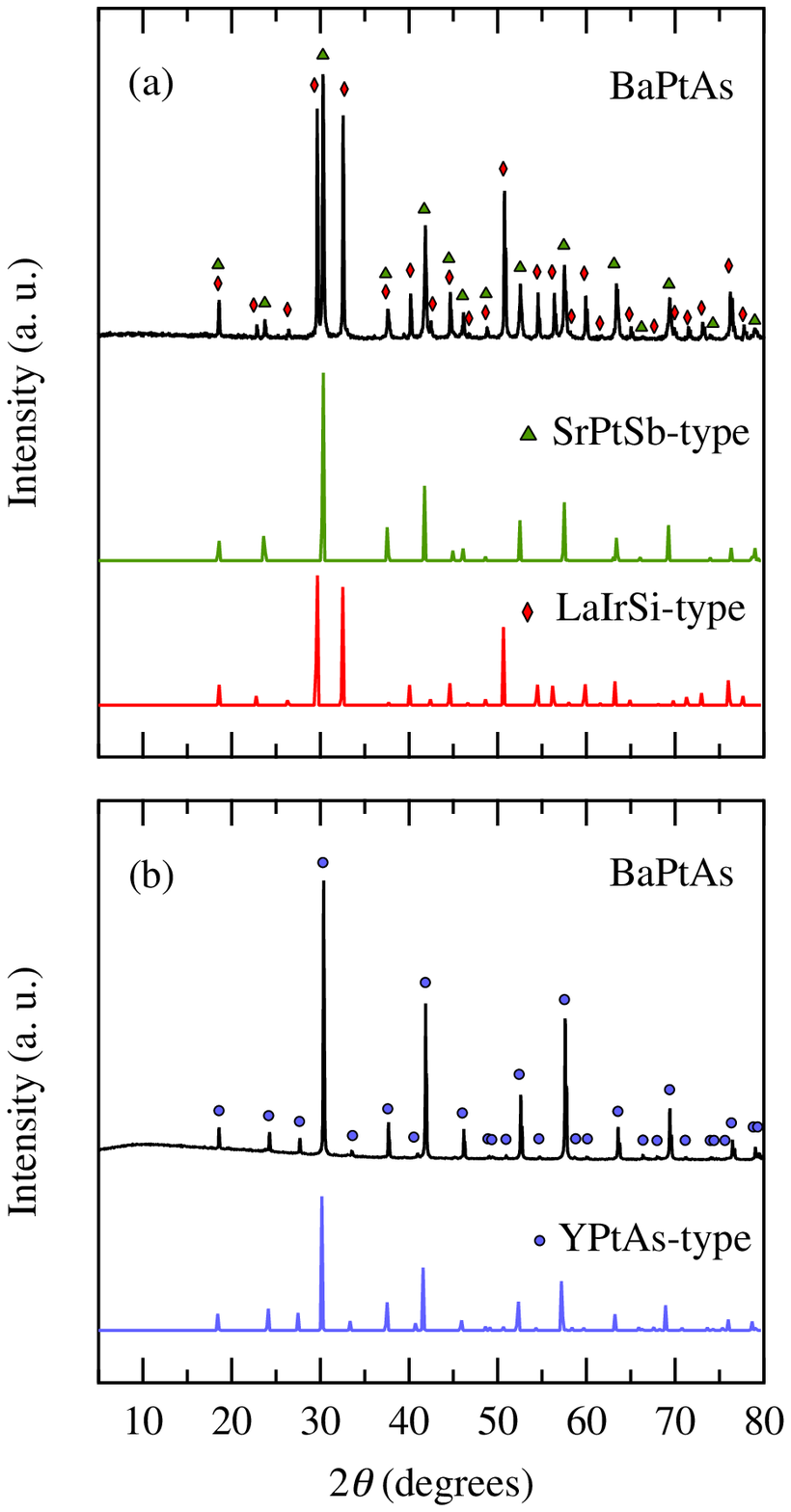}
\caption{
Powder XRD patterns for (a) the SrPtSb-type BaPtAs (space group $P\bar{6}m2$, $D_{3h}^1$, No.~187) obtained as a mixture with the LaIrSi-type  BaPtAs ($P2_13$, $T^4$, No. 198) and (b) the YPtAs-type ($P6_3/mmc$, $D_{6h}^4$, No.~194) BaPtAs. 
The green and blue lines indicate the calculated profiles of the SrPtSb- and YPtAs-type BaPtAs, respectively, based on the crystallographic parameters shown in Tables I and II. The red line indicates that of the LaIrSi-type BaPtAs based on the parameters reported in Ref. \citen{Wenski_SrBa}. 
}
\end{center}
\end{figure}

\clearpage
\subsection*{\bf 3. Lattice parameters $a$ and $c$ vs superconducting transition temperature $T_{\rm c}$ in the YPtAs-type BaPtAs}
\vspace{0.5cm}
As shown in Fig. S3, the YPtAs-type samples with different values of  $T_{\rm c}$ possess almost identical values of lattice parameters. 
\begin{figure}[thb]
\begin{center}
\includegraphics[width=6.5cm]{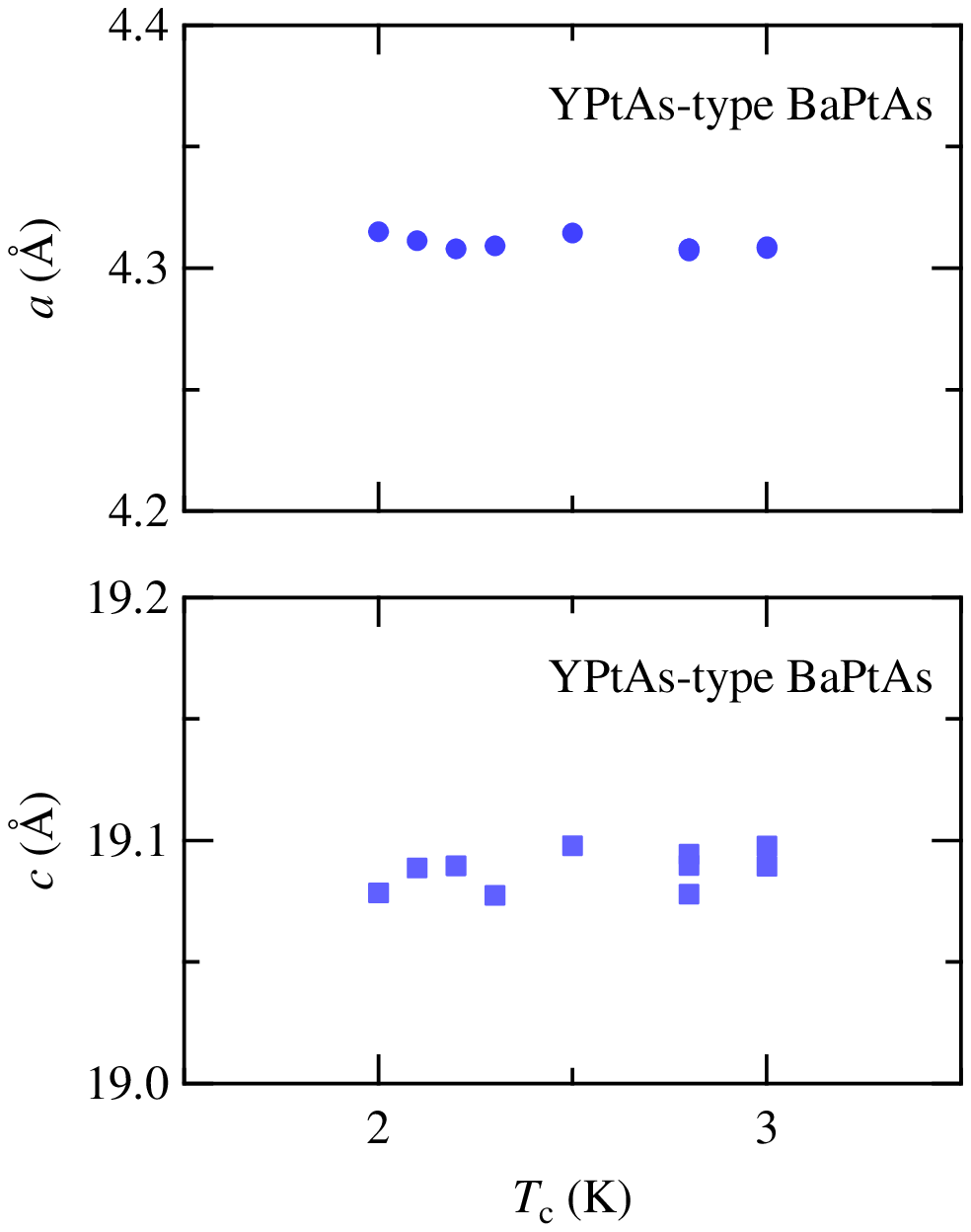}
\caption{
The plot of lattice parameters $a$ and $c$ vs superconducting transition temperature $T_{\rm c}$ for the YPtAs-type BaPtAs (space group $P6_3/mmc$, $D_{6h}^4$, No.~194).
}
\end{center}
\end{figure}


\begin{thebibliography}{99}
\bibitem{SrPtSb} The possibilities of the AlB$_2$-type BaPt$_x$As$_{2-x}$ with a statistically distributed PtAs honeycomb network and defected SrPtSb-type BaPt$_x$(Pt$_{0.10}$As$_{0.9}$) with $0.60 \leq x \leq 0.75$\cite{Wenski_SrBa} were carefully examined, as they follow the same extinction rule as the SrPtSb-type BaPtAs with an ordered honeycomb network. 
However, the structure analyses with these assumptions did not converge within the range of realistic parameters.
\end{thebibliography}
\end{document}